\documentclass{article}
\usepackage[top=1.25in, bottom=1.25in, left=1.25in, right=1.25in]{geometry}
\usepackage[utf8]{inputenc} 
\usepackage[T1]{fontenc}    
\usepackage{hyperref}       
\usepackage{url}            
\usepackage{booktabs}       
\usepackage{amsfonts}       
\usepackage{nicefrac}       
\usepackage{microtype}      
\usepackage{graphicx}
\usepackage{bm}
\usepackage{amsmath}
\usepackage[square,numbers]{natbib}
\usepackage{subfig}
\usepackage{placeins}
\usepackage{authblk}
\usepackage[title]{appendix}

\title{\textbf{Turbulence Enrichment with Physics-informed Generative Adversarial Network}\thanks{Code for our model is available at \url{https://github.com/akshaysubr/TEGAN}}}

\author[1,3]{Akshay Subramaniam\thanks{Email: \texttt{akshays@stanford.edu}}}
\author[1,3]{Man Long Wong\thanks{Email: \texttt{wongml@stanford.edu}}}
\author[1]{Raunak D. Borker}
\author[1]{Sravya Nimmagadda}
\author[1,2,3]{Sanjiva K. Lele}
\affil[1]{Department of Aeronautics \& Astronautics, Stanford University}
\affil[2]{Department of Mechanical Engineering, Stanford University}
\affil[3]{Center for Turbulence Research, Stanford University}
\date{}                     

\begin{document}

\maketitle

\begin{abstract}
Generative Adversarial Networks (GANs) have been widely used for generating photo-realistic images \cite{goodfellow2014generative,radford2015unsupervised,gulrajani2017improved}. A variant of GANs called super-resolution GAN (SRGAN) \cite{ledig2016photo} has already been used successfully for image super-resolution where low resolution images can be upsampled to a $4\times$ larger image that is perceptually more realistic. However, when such generative models are used for data describing physical processes, there are additional known constraints that models must satisfy including governing equations and boundary conditions. In general, these constraints may not be obeyed by the generated data. In this work, we develop physics-based methods for generative enrichment of turbulence. We incorporate a physics-informed learning approach by a modification to the loss function to minimize the residuals of the governing equations for the generated data. We have analyzed two trained physics-informed models: a supervised model based on convolutional neural networks (CNN) and a generative model based on SRGAN: Turbulence Enrichment GAN (TEGAN), and show that they both outperform simple bicubic interpolation in turbulence enrichment. We have also shown that using the physics-informed learning can also significantly improve the model's ability in generating data that satisfies the physical governing equations. Finally, we compare the enriched data from TEGAN to show that it is able to recover statistical metrics of the flow field including energy metrics and well as inter-scale energy dynamics and flow morphology.
\end{abstract}

\section{Introduction}
Predicting turbulence accurately is extremely challenging especially in capturing high-order statistics due to its intermittent nature. Although it is well-known that turbulence can be described very precisely with the three-dimensional (3D) Navier--Stokes equations, so far no known general analytical solution exists for the equations. While numerical methods provide another path for attainable approximate solutions, the computational resources required for direct numerical simulations (DNSs) with Navier--Stokes equations for engineering applications can be quite large, especially when the flows are highly turbulent, or the Reynolds numbers are high, where a broad range of spatial and temporal scales appears. As a result, reduced-order modeling of turbulent flows for low cost computations has been a popular research area in fluid dynamics for decades.

Traditional turbulence modeling requires the mathematical modeling of Reynolds stress tensor in Reynolds-averaged Navier--Stokes (RANS) equations for RANS simulations, or subgrid/subfilter scale effects in filtered Navier--Stokes equations for large-eddy simulations (LESs). Although high-fidelity simulation and experimental data are often used to validate these models, it is still not guaranteed that they are in the best functional forms since they are usually introduced based on only physical intuitions and heuristic arguments. Machine learning (ML), on the other hand, may provide an alternative efficient strategy towards constructing an optimal mapping between available turbulence data and statistical quantities of interest. Despite the pioneering work on using neural networks to reconstruct turbulent fields near wall by~\citet{milano2002neural}, the use of ML for turbulence modeling has not received much attention until recent years~\cite{duraisamy2019turbulence,brunton2019machine} after successful applications of deep neural networks (DNNs) in different fields, such as health care, computer vision, speech recognition, sales forecasting, etc, are seen. \citet{zhang2015machine} investigated the use of neural networks to reconstruct the spatial form of correction factors for the turbulent and transitional flows. \citet{ling2016reynolds} developed a model for the Reynolds stress anisotropy tensor using DNN with embedded Galiliean invariance and the accuracy of their DNN over two conventional RANS models was demonstrated by \citet{kutz2017deep}. A synthetic turbulence generator for inflow using a convolutional neural network (CNN) is developed by~\citet{fukami2019synthetic}. \citet{lapeyre2019training} demonstrated an approach with CNN to model subgrid scale reaction rates.

Traditional LES approach that only models the subgrid scale effects on the filtered velocity field may not be sufficient for applications that need more refined data that is not available on the LES grid. For example, in particle-laden flow LESs, the full-scale carrier-phase velocity is required to accurately describe the transport of particles inside the flows~\cite{bassenne2019dynamic}. In LESs of wind farm flows, the prediction of fatigue loading in wind turbines that operate within the planetary boundary layer needs more refined representation of the subgrid velocity fluctuations that is statistically accurate~\cite{ghate2017subfilter}. Recently, \citet{fukami2019super} analyzed the use of ML models including deep CNN for super-resolution of chaotic flows. While their methods may serve the purpose of turbulence enrichment, they only showed capabilities of their models for two-dimensional (2D) ``turbulent'' flows. Since the dynamics of turbulence is greatly affected by 3D vortex-stretching effect, further investigation is necessary to determine whether the state-of-the-art DNNs can be used for turbulence enrichment under physical constraints.

The CNN used in the work~\cite{fukami2019super} is based on the super-resolution CNN (SRCNN) by~\citet{dong2015image} which was developed mainly for the high resolution reconstruction of low resolution images using deep CNN. This seminal work shows superior accuracy of the proposed model compared with other state-of-the-art image reconstruction methods. Since then, there have been many works~\cite{dong2016accelerating,lim2017enhanced,tai2017image,lai2017deep} on improving DNNs for super-resolution of images. However, those models still cannot handle large up-scaling factor and generate images with satisfying high-wavenumber details. In the work by~\citet{ledig2016photo}, the aforementioned problems were tackled with the use of a generative adversarial network (GAN) for super-resolution of images. The idea of GAN, which is composed of two neural networks (generator and discriminator) that contest with each other, was invented by~\citet{goodfellow2014generative}. In a GAN, the generator creates outputs while the discriminator evaluates them.
GANs have been shown to perform better than other data driven approaches like PCA in capturing high-order moments~\cite{chan2017parametrization}, and thus may be beneficial for turbulence modeling. While one may attempt to directly extend the super-resolution GAN (SRGAN)~\cite{ledig2016photo} for turbulence enrichment of 3D low resolution turbulent data, the generated high resolution data may not be physically realistic as physical constraints are not present when these DNNs are designed for image upsampling. Incorporating the physical constraints into DNNs such as the SRGAN framework would be crucial to its performance in this context.

In this work, we propose the extensions of the CNN-based residual block neural network~\cite{he2016deep} (TEResNet) and the SRGAN~\cite{ledig2016photo} (TEGAN) to enrich low resolution turbulent fields with high wavenumber contents using physics-informed technique. The input to our model is a low resolution turbulent flow field that consists of four 3D fields each of size $16 \times 16 \times 16$. We then apply both TEResNet and TEGAN to upsample each of these four fields to $64 \times 64 \times 64$. To our knowledge physics-informed CNNs or GANs for turbulence enrichment have not been studied previously. In the works~\cite{raissi2017physics,raissi2017physics2}, physics-based neural networks were developed to infer solutions to a partial differential equation (PDE). To achieve this they used a mean squared error loss function with contributions from not only the error in the solution, but also from the residual of the governing PDE. This physics-informed technique was implemented and tested on a one-dimensional (1D) model problem, but may not be a feasible approach for solving large 3D chaotic systems like turbulent flows. Our results show that the proposed TEResNet and TEGAN models have better performance than only using tricubic interpolation for upsampling and the turbulence enriched data compares well with the high resolution data.


\section{Problem description and governing equations}
The demonstration problem that we have chosen is the incompressible forced isotropic homogeneous turbulence problem in a 3D ($x$-$y$-$z$ or $x_1$-$x_2$-$x_3$) periodic domain with size $[0, 2\pi)\times[0, 2\pi)\times[0, 2\pi)$. The governing equations are the time ($t$) dependent incompressible Navier--Stokes equations:
\begin{align}
    \nabla \cdot \bm{u} &= 0, \label{eq:contunity} \\
    \frac{\partial \bm{u}}{\partial t} + \bm{u} \cdot \nabla \bm{u} &= \nabla p + \nu \nabla^2 \bm{u}, \label{eq:momentum}
\end{align}
where $\bm{u}=\left[u, v, w \right]^T=\left[u_1, u_2, u_3 \right]^T$, $p$, and $\nu$ are the velocity vector, kinematic pressure, and kinematic shear viscosity, respectively. Equation~\eqref{eq:contunity} is the continuity equation derived from the conservative of mass and Equation~\eqref{eq:momentum} is the conservation of momentum. Taking the divergence of the momentum equation and using the continuity equation, we can get a Poisson equation for pressure:
\begin{equation}
    - \nabla^2 p = \nabla \bm{u} : \nabla \bm{u}^T .
\end{equation}
The continuity equation and the pressure Possion equation can be viewed as physical constraints on the velocity and pressure fields respectively.



\section{Dataset and features}

\subsection{Data generation}

Data is generated using an in-house computational fluid dynamics (CFD) simulation code: PadeOps. Spectral methods are used for the spatial discretizations and a third-order total variation diminishing Runge--Kutta (RK-TVD) scheme~\cite{shu1989efficient} is adopted for
time integration.
We perform a time-evolving DNS on a $64\times64\times64$ uniform grid and collect snapshots separated in time by more than one integral time scale. This ensures that each example is statistically decorrelated. Each snapshot is comprised of four fields - three components of the velocity vector $( u, v, w )$ and the kinematic pressure, $(p)$, each of size $64 \times 64 \times 64$. Low resolution data is then generated by filtering the high resolution data down to $16 \times 16 \times 16$ using a compact support explicit filter discussed in Appendix~\ref{app:low_pass_filter}. The filter is derived as an approximation to the sharp spectral low-pass filter at cutoff of a quarter of Nyquist wavenumber with a compact stencil. The velocity components of the high resolution data are normalized (and non-dimensionalized) by the root mean square of the velocity magnitude and the pressure by the mean square of the velocity magnitude. The dataset is divided as Train/Dev/Test split: $920\ (79.3\%)/120\ (10.3\%)/120\ (10.3\%)$. Sample images of the high and low resolution data are presented in Section~\ref{sec:results}.

\section{Methods}

\subsection{Architecture of different neural network models }
For the task of upsampling the low resolution data in a physically consistent manner, we use a GAN\cite{goodfellow2014generative} in a fashion similar to super-resolution applications for image data \cite{ledig2016photo}.

\begin{figure}
\centering
\includegraphics[width=0.99\textwidth]{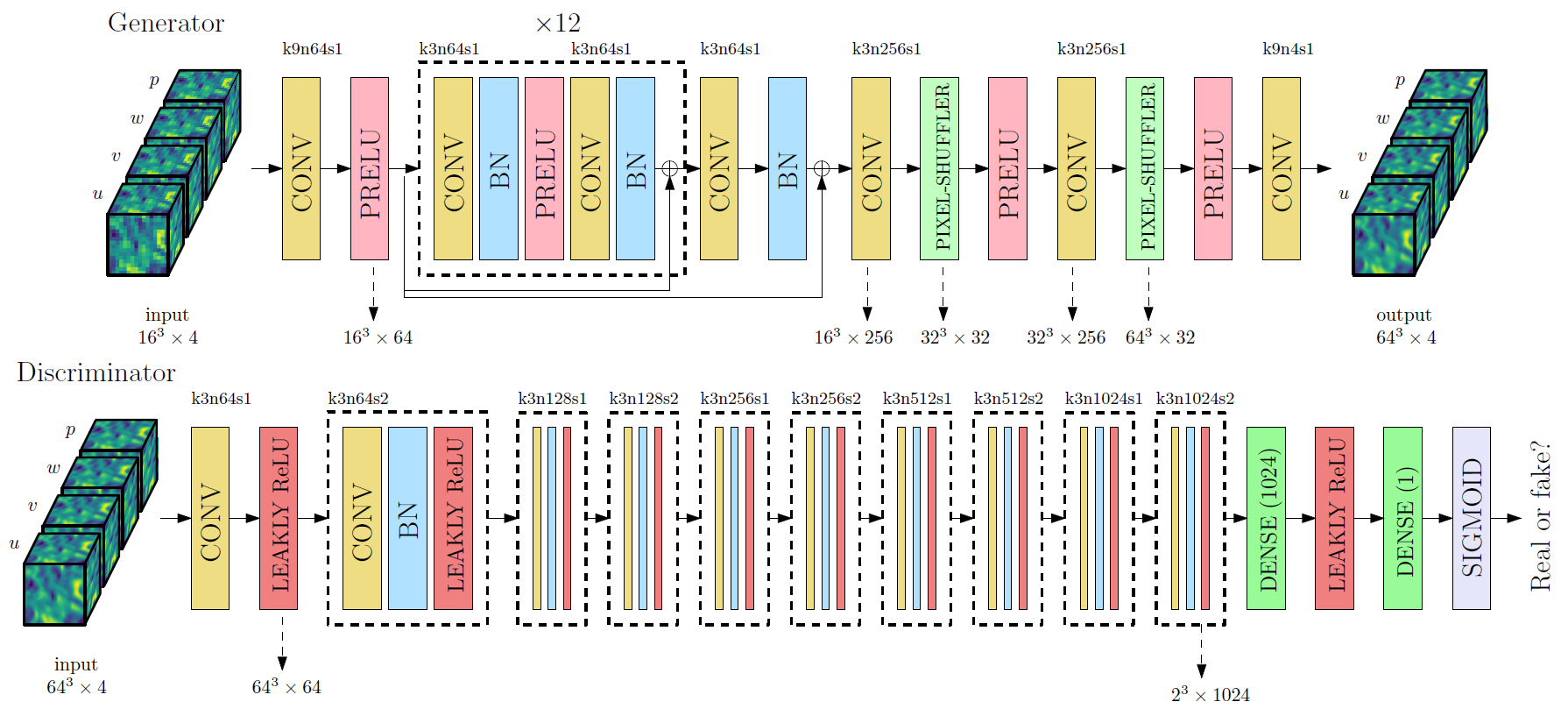}
\caption{Architecture of generator and discriminator network with corresponding kernel size (k), number of channels (n) and stride (s) indicated for each convolutional layer.}
\label{TEGAN_arch}
\end{figure}

The generator has a deep residual network architecture with each residual block having convolutional layers with batch normalization. The discriminator has a deep convolutional architecture with fully connected layers in the end for binary classification. The architectures of the generator and discriminator are depicted pictorially in Figure~\ref{TEGAN_arch}.

\subsection {Loss functions}
As discussed in a previous section, the flow field is constrained by the continuity and pressure Poisson equations which are repeated below:
\begin{align*}
\nabla \cdot \bm{u} &= 0, \\
- \nabla^2 p &= \nabla \bm{u} : \nabla \bm{u}^T.
\end{align*}
The above equations might not be satisfied exactly by the model's generated output. To counter this, the residual of the above equations can be used as a regularizer for the model through a physics loss.

The loss function minimized for the generator network during training is a combination of a content loss $\mathcal{L}_\mathrm{content}$ and a physics loss $\mathcal{L}_\mathrm{physics}$.

\begin{align}
\mathcal{L}_\mathrm{GAN} &= \left( 1 - \lambda_\mathrm{A} \right) \mathcal{L}_\mathrm{resnet} + \lambda_\mathrm{A} \mathcal{L}_\mathrm{adversarial}, \\
\mathcal{L}_\mathrm{resnet} &= \left( 1 - \lambda_\mathrm{P} \right) \mathcal{L}_\mathrm{content} + \lambda_\mathrm{P} \mathcal{L}_\mathrm{physics}, \\
\mathcal{L}_\mathrm{content} &= \left( 1 - \lambda_\mathrm{E} \right) \mathcal{L}_\mathrm{MSE} + \lambda_\mathrm{E} \mathcal{L}_\mathrm{enstrophy}, \\
\mathcal{L}_\mathrm{physics} &= \left( 1 - \lambda_\mathrm{C} \right) \mathcal{L}_\mathrm{pressure} + \lambda_\mathrm{C} \mathcal{L}_\mathrm{continuity} .
\end{align}

\begin{figure}
    \centering
    \includegraphics[width=0.5\textwidth]{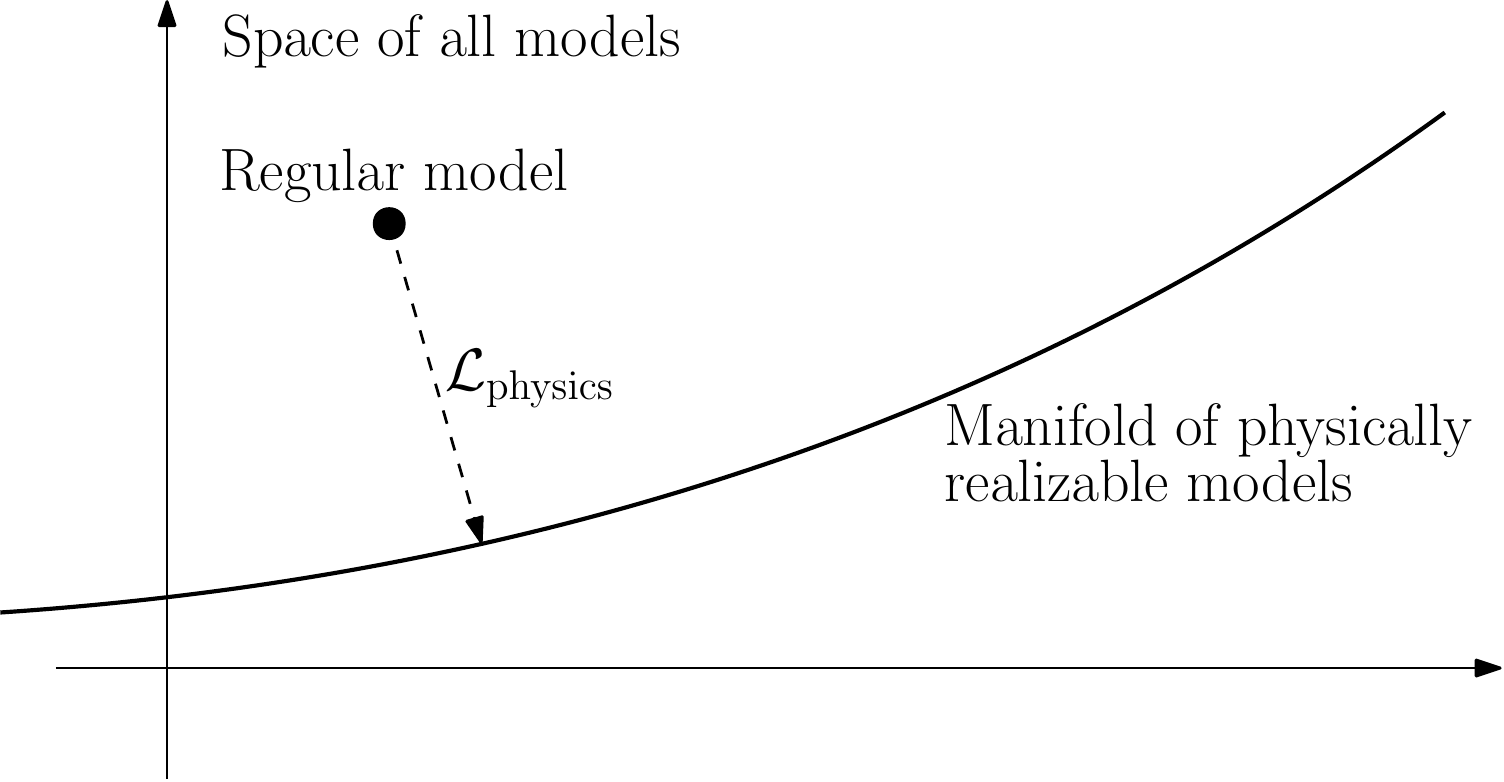}
    \caption{Explanation of the role of physics loss.}
    \label{fig:physics_loss_schematic}
\end{figure}

\begin{itemize}
\item \textbf{Content loss}: $\mathcal{L}_\mathrm{content}$

$\mathcal{L}_\mathrm{MSE}$: Mean squared error between the high resolution and generated fields.

$\mathcal{L}_\mathrm{enstrophy}$: Mean squared error in the derived enstrophy field $\Omega$ ($\Omega = \bm{\omega} \cdot \bm{\omega},\ \mathrm{where}\ \bm{\omega}=\nabla \times \bm{u}$) to sensitize the generator to high frequency content.

\item \textbf{Physics loss}: $\mathcal{L}_\mathrm{physics}$
Residuals of the continuity ($\mathcal{L}_\mathrm{continuity}$) and pressure Poisson ($\mathcal{L}_\mathrm{pressure}$) equations given above similar to \cite{raissi2017physics}. As depicted in figure above, the inclusion of physics loss forces the our NN to generate on physically realizable solutions. 

\item \textbf{Adversarial loss}: a logistic loss function is used similar to that defined in \citep{ledig2016photo}.
\end{itemize}
To train the discriminator, we use the logistic loss based on predicted labels for real and generated data.

\subsection{Training}
\subsubsection{TEResNet}
A model with just the residual generator network without the adversarial component is termed TEResNet. We first train TEResNet to convergence and tune hyperparameters like the number of residual blocks and the physics loss parameters.
\subsubsection{TEGAN}
The model with both the residual network generator and the discriminator depicted above is termed TEGAN. The generator in TEGAN is first initialized using the weights from the trained TEResNet while the discriminator is initialized using the Xavier--He initialization~\cite{glorot2010understanding,he2015delving}.

For the first few iterations in the training process ($\sim 300$), the discriminator alone is trained to negate the advantage that the generator has because of its pre-trained weights. Then, both the generator and discriminator are trained together with the active adversarial loss until the losses saturate and the discriminator's output saturates at 0.5. The Adam optimizer~\cite{kingma2014adam} is used for updating the weights and training both the networks.

\section{Results}
\label{sec:results}

\subsection{Training convergence}
\begin{figure}[ht!]
\centering
\includegraphics[width=0.49\textwidth]{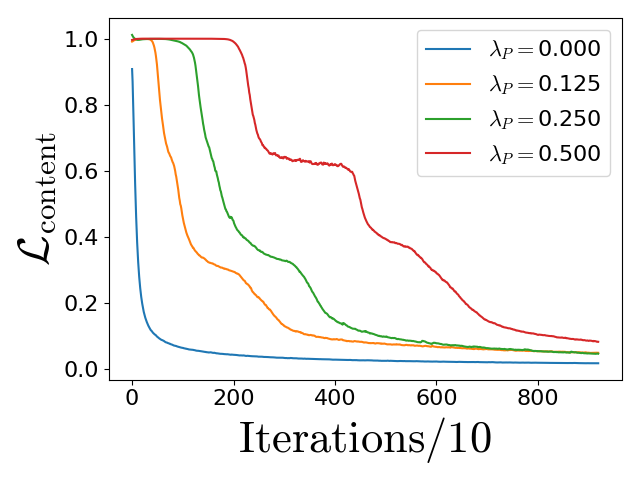}
\includegraphics[width=0.49\textwidth]{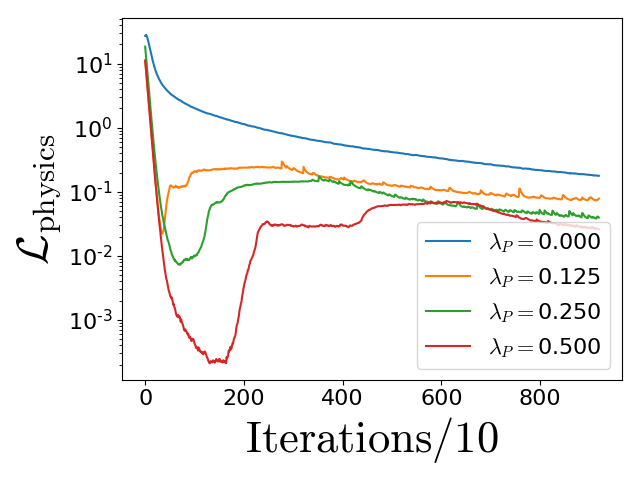}
\caption{Losses against iterations. Left: content loss; right: physics loss.}
\label{fig:losses}
\end{figure}

A batch size of 5 was chosen for training of both TEResNet and TEGAN because of memory constraints of the GPU used for training. We choose $\alpha=1.0\mathrm{e}{-4}$ as the learning rate for training from the hyper-parameter search. We also examine the effect of the physics loss weight $\lambda_\mathrm{P}$ through experiments on TEResNet. Figure \ref{fig:losses} shows the content and physics losses during training of TEResNet with different values for $\lambda_{\mathrm{P}}$.We see that adding a non-zero weight to the physics loss improves the physics residual by almost an order of magnitude. We choose $\lambda_{\mathrm{P}}=0.125$ as this gives good compromise between the two losses. Another interesting observation is that for higher weightage to the physics loss, the trivial solution of zero fields becomes a local minimum.

\begin{figure}[ht!]
  \centering
    \includegraphics[width=0.49\textwidth]{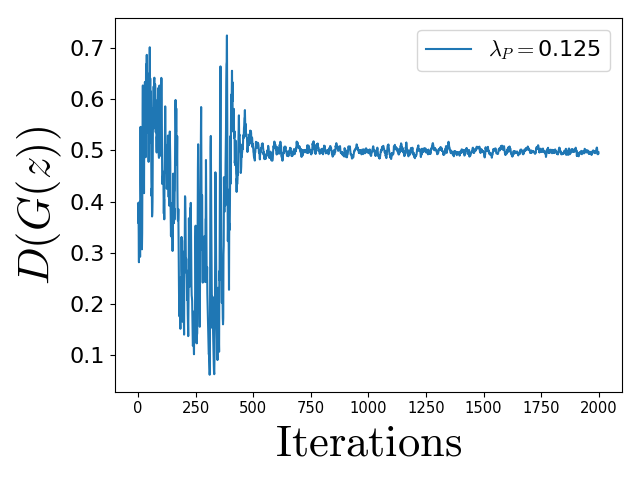}
    \includegraphics[width=0.49\textwidth]{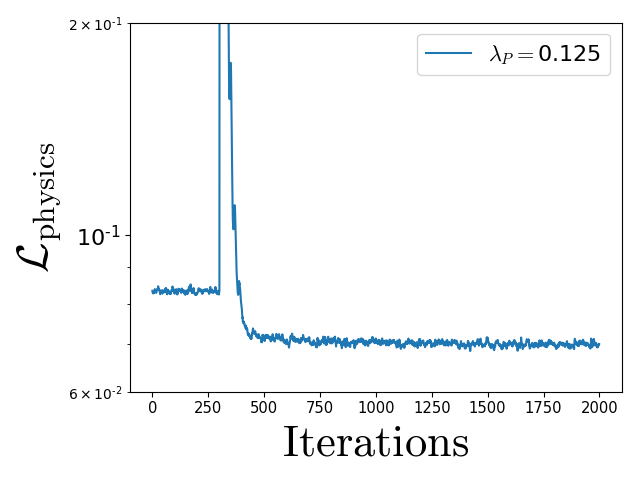}
    \caption{Left: discriminator output for generated data; right: physics loss of TEGAN.}
    \label{fig:discrimin}
\end{figure}

To train the TEGAN model, we initialize its weights using the trained TEResNet for the generator. We then train only the discriminator until the discriminator is able to distinguish between real and generated data. Then we train the discriminator and generator together training the discriminator twice as often as the generator. To improve the stability for training TEGAN, we add a staircase decay for the learning rate. We set the decay rate to $0.5$ and chose a decay step of $400$ by running a case without learning rate decay and estimating the number of steps required to go close to the minimum. Figure \ref{fig:discrimin} shows the convergence of TEGAN during training. It can be seen that the discriminator output for generated data saturates at 0.5 and the physics loss converges to a smaller value compared to the initial value from TEResNet.

\subsection{Model evaluation}
\begin{figure}[ht!]
    \centering
        \subfloat{\includegraphics[width=0.99\textwidth]{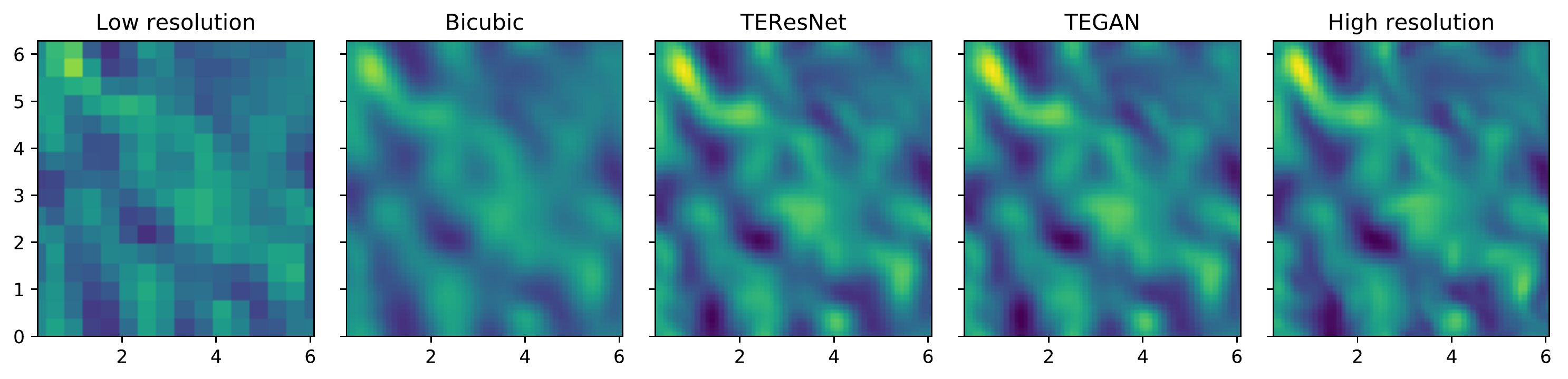}} \\
        \subfloat{\includegraphics[width=0.99\textwidth]{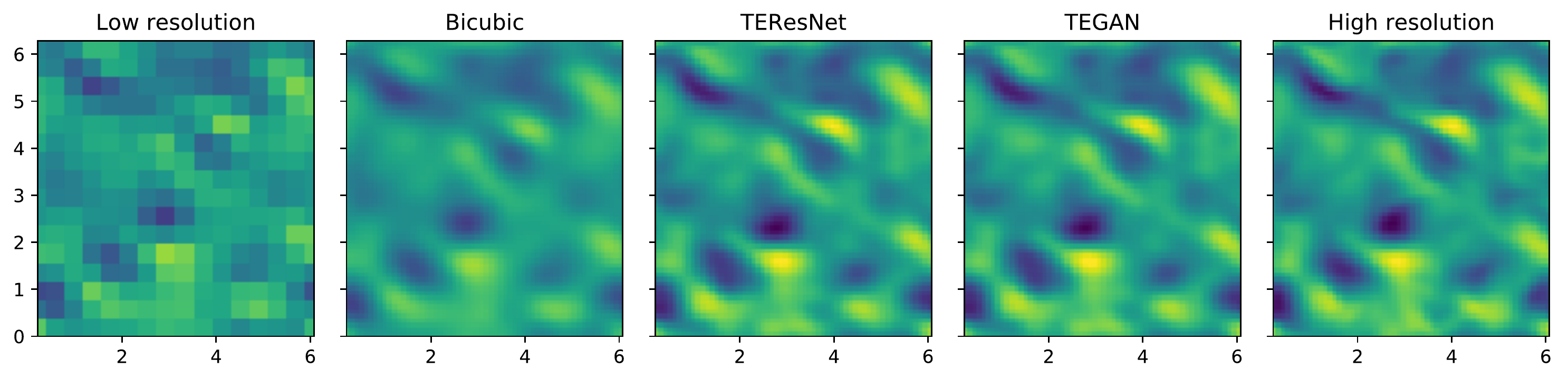}} \\
        \subfloat{\includegraphics[width=0.99\textwidth]{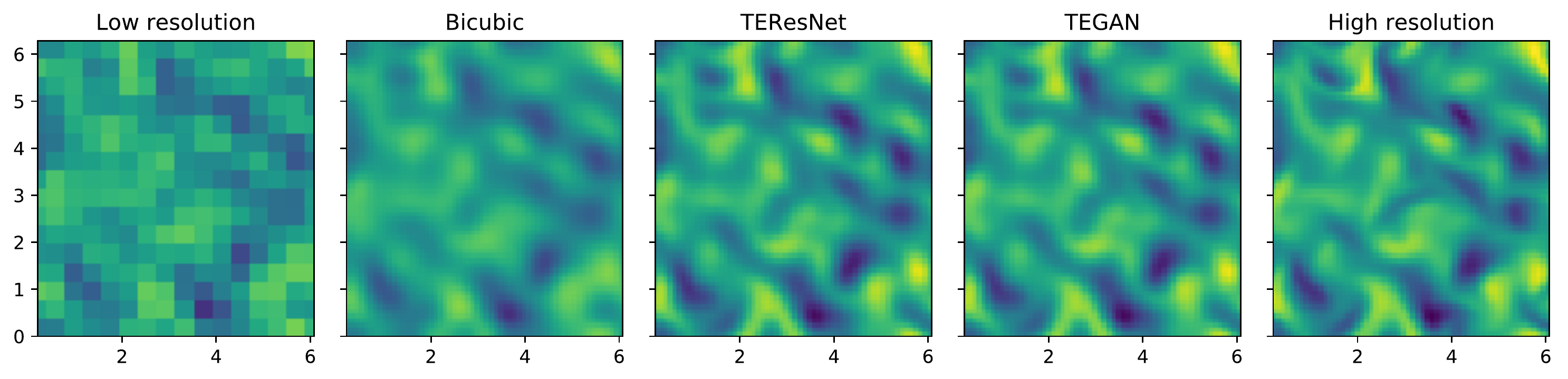}} \\
        \subfloat{\includegraphics[width=0.99\textwidth]{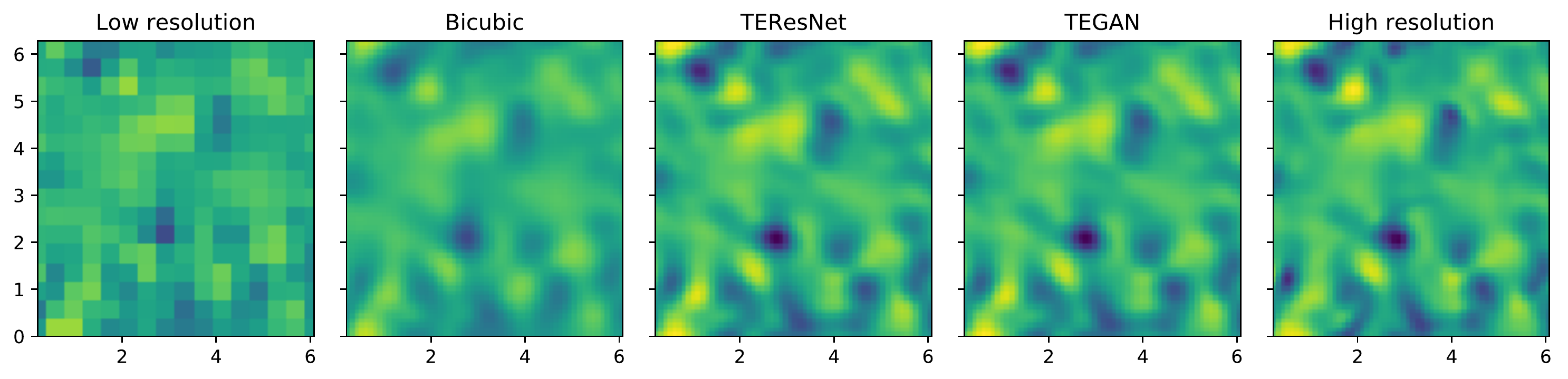}}
    \caption{Comparisons of the (from left to right) low resolution, tricubic interpolation, TEResNet TEGAN and high resolution fields. Plots are of the $u$, $v$, $w$ and $p$ (from top to bottom) on a slice of the 3D field.}
    \label{fig:comparison}
\end{figure}

\begin{figure}[ht!]
  \centering
    \includegraphics[width=0.6\textwidth]{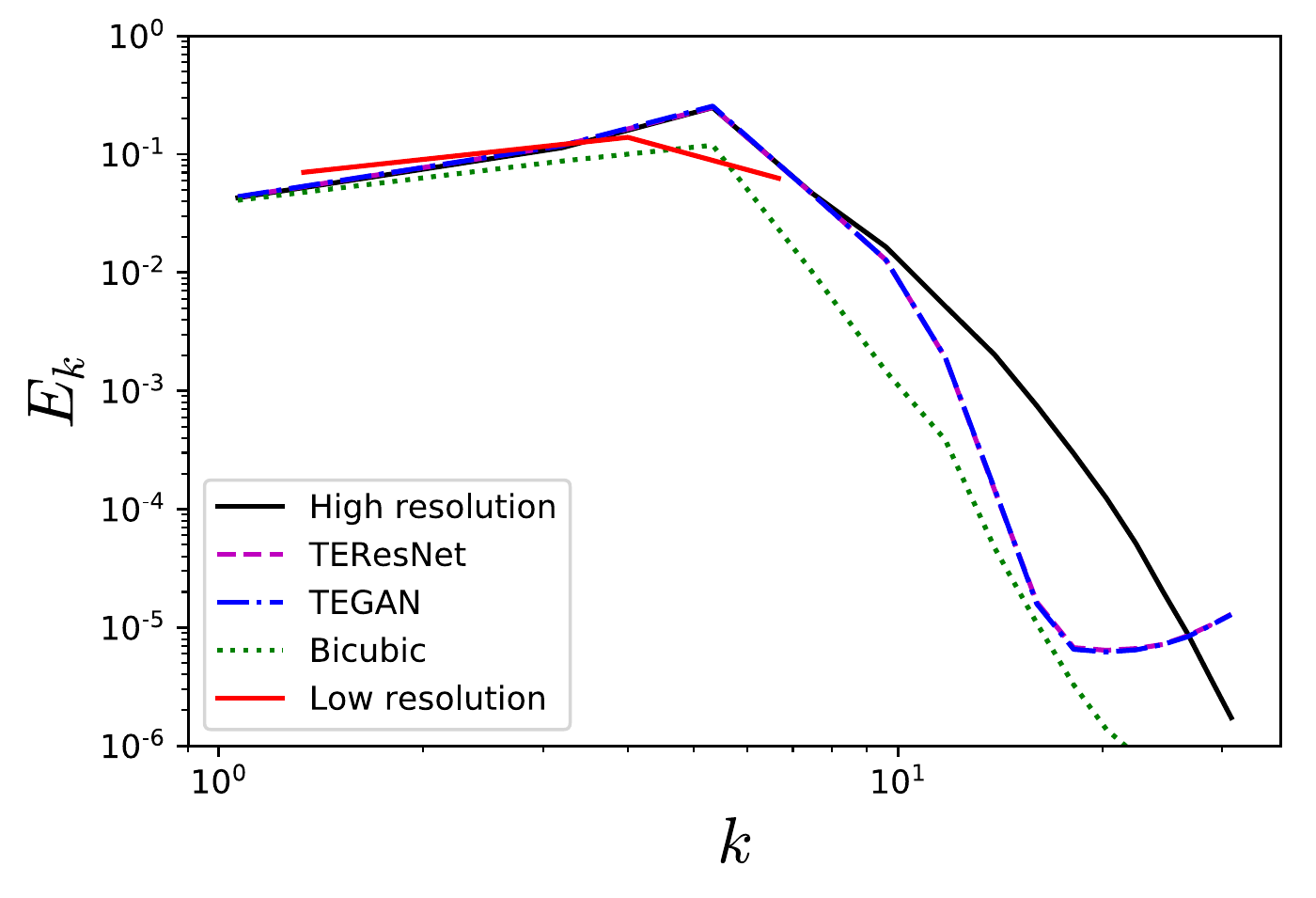}
    \caption{Comparison of the velocity energy spectra for the different upsampling methods.}
    \label{fig:spectra}
\end{figure}

Figure~\ref{fig:comparison} compares the qualities of upsampled data from tricubic interpolation, TEResNet and TEGAN. Both TEResNet and TEGAN outperform the tricubic interpolation in reconstructing small-scale features. This is also evident from the plots of the velocity energy spectra in Figure~\ref{fig:spectra}. The output from TEResNet and TEGAN are indistinguishable visually but Table~\ref{tab:comparison} shows that there is more than $10\%$ improvement of TEGAN over TEResNet in minimizing the physics loss while the content losses of both models are similar.

\begin{table}[ht!]
\centering
\label{my-label}
\begin{tabular}{c|c|c|c|c|}
\cline{2-5}
\multicolumn{1}{l|}{} & \multicolumn{2}{c|}{\textbf{\begin{tabular}[c]{@{}c@{}}$\mathcal{L}_\mathrm{content}$\end{tabular}}} & \multicolumn{2}{c|}{\textbf{\begin{tabular}[c]{@{}c@{}}$\mathcal{L}_\mathrm{physics}$\end{tabular}}} \\ \cline{2-5} 
 & Dev & Test & Dev & Test \\ \hline
\multicolumn{1}{|c|}{\textbf{TEResNet}} & 0.049 & 0.050 & 0.078 & 0.085 \\ \hline
\multicolumn{1}{|c|}{\textbf{TEGAN}} & 0.047 & 0.047 & 0.070 & 0.072 \\ \hline \hline
\multicolumn{1}{|c|}{\textbf{\%\ Difference}} & 4.1 & 6.0 & 10.3 & 15.2 \\ \hline
\end{tabular}
\caption{Comparison of losses between TEResNet and TEGAN at the end of training. \label{tab:comparison}}
\end{table}

\subsection{Higher order correlations}
\label{sec:HOC}


The fundamental statistical theory of homogeneous turbulence arises from the analysis of second order velocity structure function $D_{ij} \left(\bm{r}, t \right)$ given by
\begin{equation}
    D_{ij} \left(\bm{r}, t \right) = \langle \left[ u_i(\bm{x}, t) - u_i(\bm{x}+\bm{r}, t)\right] \left[ u_j(\bm{x}, t) - u_j(\bm{x}+\bm{r}, t)\right] \rangle ,
\end{equation}
where $\bm{r}$ is the separation vector, $u_i\left(\bm{x},t\right)$ is the velocity in the $x_i$ direction and $\langle \cdot \rangle$ denotes statistical averaging.
In homogeneous isotropic turbulence, the second order velocity structure function depends only on two second order tensors as
\begin{equation}
    D_{ij} \left(\bm{r}, t \right) = D_{NN} \left(r, t \right)\delta_{ij} + \left[ D_{LL} \left(r, t \right) - D_{NN} \left(r, t \right) \right]\frac{r_ir_j}{r^2}
\end{equation}
where $D_{LL}$ and $D_{NN}$ are the longitudinal and transverse structure functions respectively.
Further, in homogeneous isotropic turbulence, the second order structure function can be related to the two-point correlation tensor $R_{ij}\left(\bm{r}\right)$ as
\begin{equation}
    D_{ij} \left(\bm{r}, t \right) = 2 R_{ij} \left(\bm{0}, t \right) - R_{ij} \left(\bm{r}, t \right) - R_{ji} \left(\bm{r}, t \right).
\end{equation}
Therefore, the second order structure functions in homogeneous isotropic turbulence are fully characterized by the longitudinal and transverse two-point correlation functions $R_{11} \left(r\bm{e}_1, t \right)$ and $R_{22} \left(r\bm{e}_1, t \right)$ respectively \cite{monin1971statistical}. Here, $\bm{e}_1$ represents the unit vector along the $x$ direction.

\begin{figure}
    \centering
        \subfloat{\includegraphics[width=0.5\textwidth]{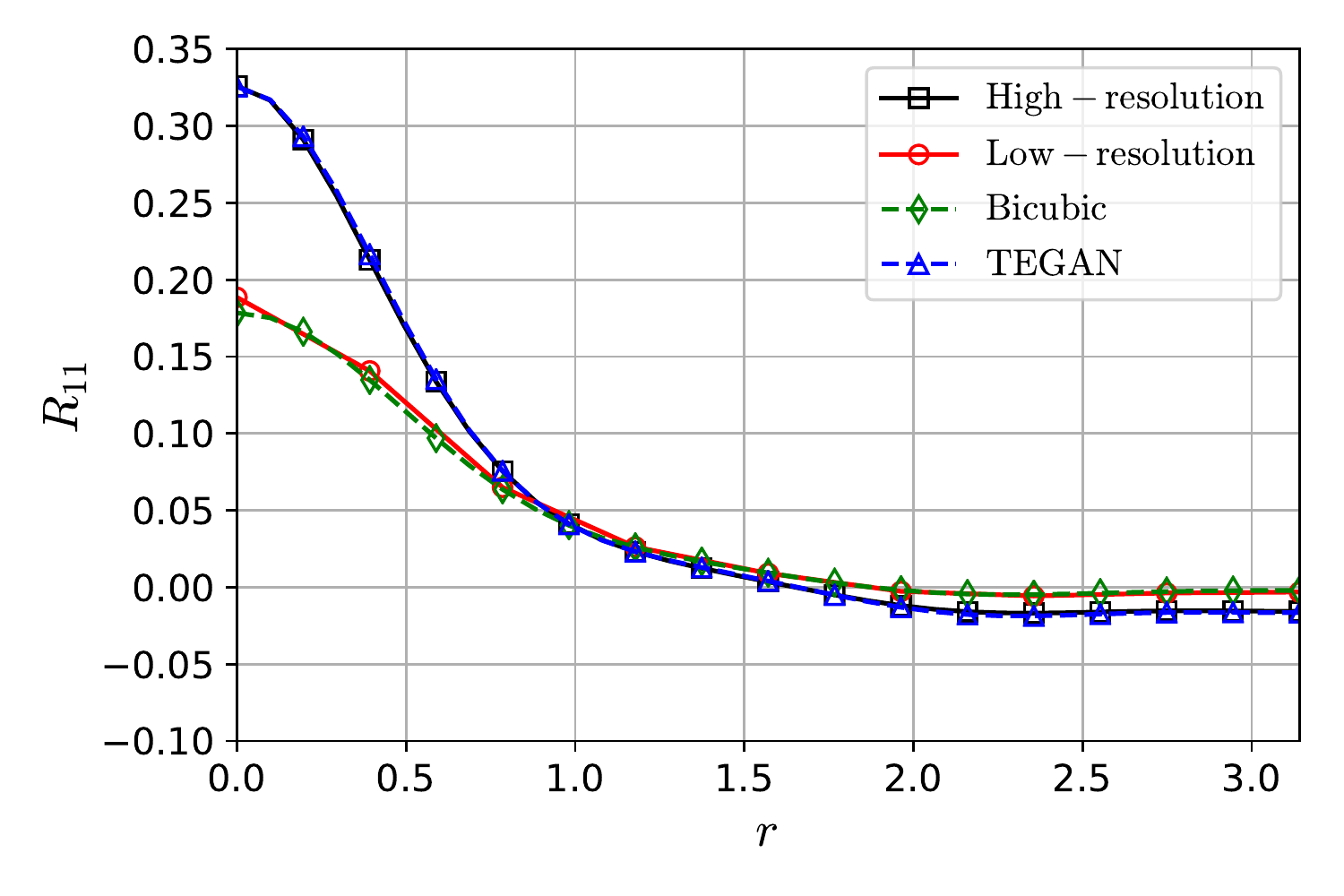}}
        \subfloat{\includegraphics[width=0.5\textwidth]{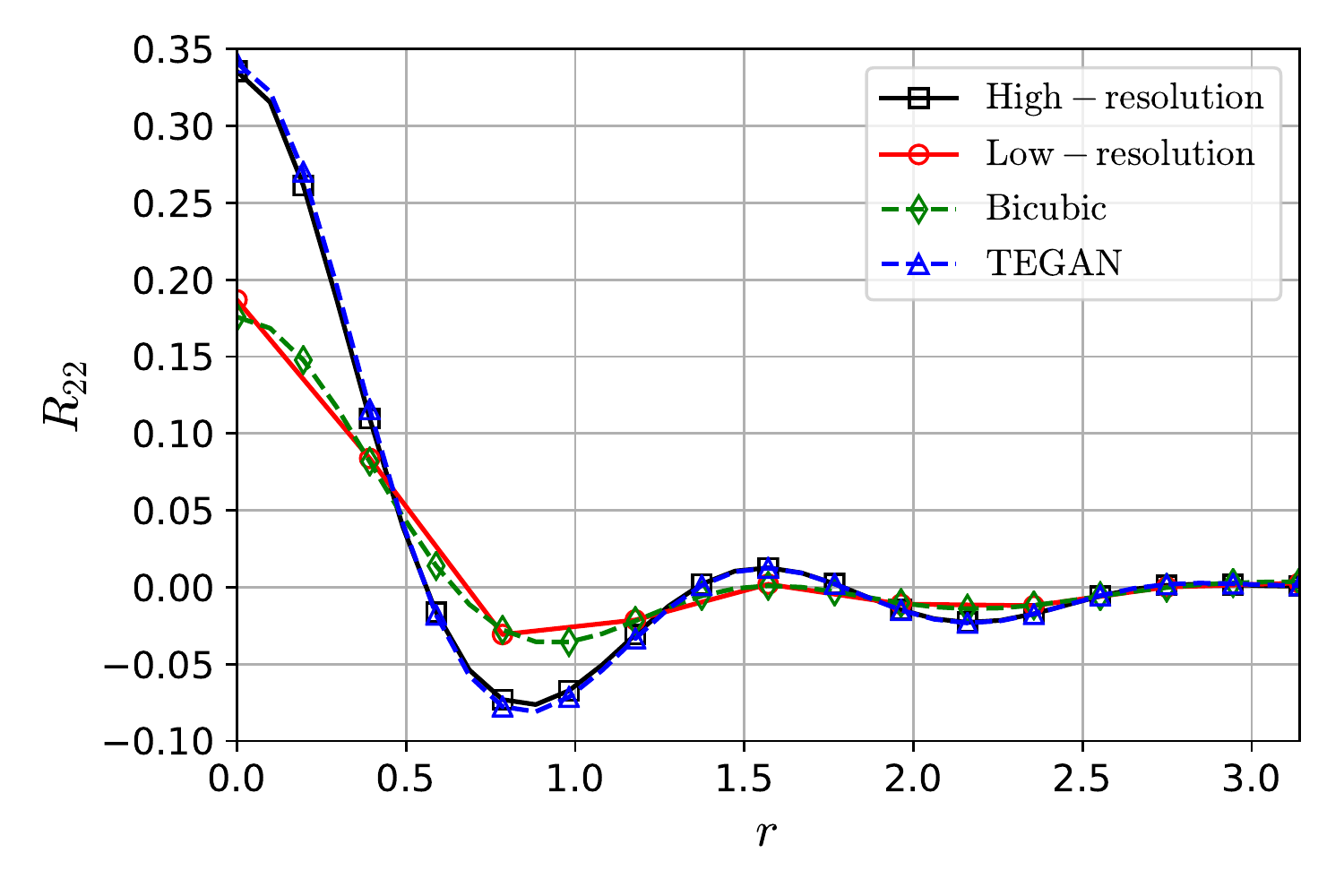}}
    \caption{Longitudinal (left) and transverse (right) two-point correlations compared from the low resolution data (red circles), high resolution data (black squares), bicubic interpolation (green diamonds) and TEGAN (blue triangles). \label{fig:TPC_R} }
\end{figure}

Figure~\ref{fig:TPC_R} shows the longitudinal and transverse two-point correlation functions for the low and high resolution datasets as well as for bicubic interpolation and the data generated by the TEGAN model. Since the two-point correlation at zero separation ($r = 0$) is the Turbulent Kinetic Energy (TKE), the low resolution has a much lower value than the high resolution data at $r = 0$. This is to be expected since the finer energy containing scales are filtered out in the low resolution data. Bicubic interpolation basically acts to interpolate the low resolution two point correlations and completely misrepresents the true correlations. The TEGAN model however is able to faithfully recover both longitudinal and transverse two-point correlation functions and is very close to the true high resolution values. This shows that the TEGAN model recovers the energy in the fine scales as well as accurately represents the statistical properties of the turbulent field. Since the two-point correlation functions are recovered accurately, this also means that derived statistical quantities like the integral length scale are also accurately represented.

\begin{figure}
    \centering
        \subfloat{\includegraphics[width=0.5\textwidth]{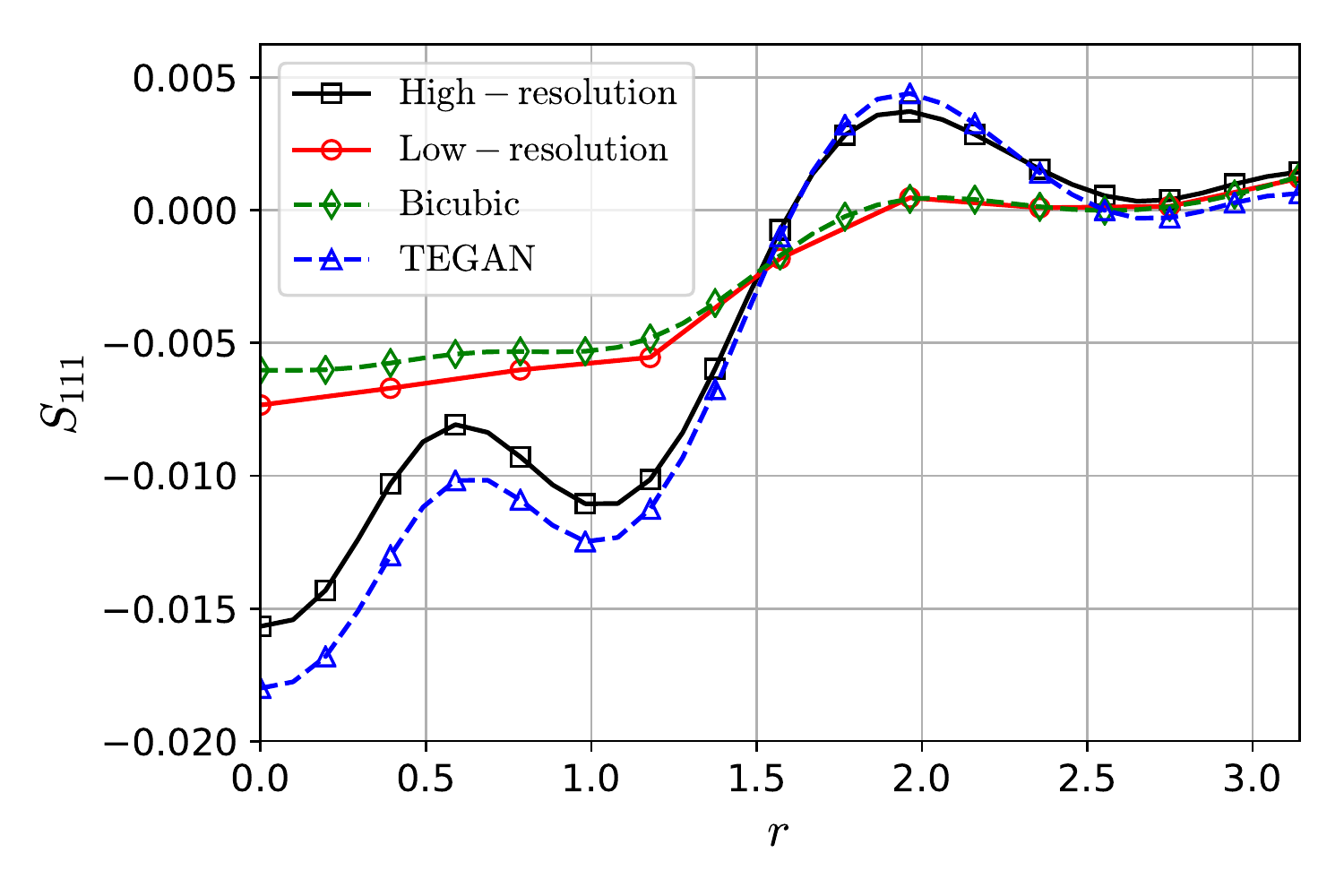}}
        \subfloat{\includegraphics[width=0.5\textwidth]{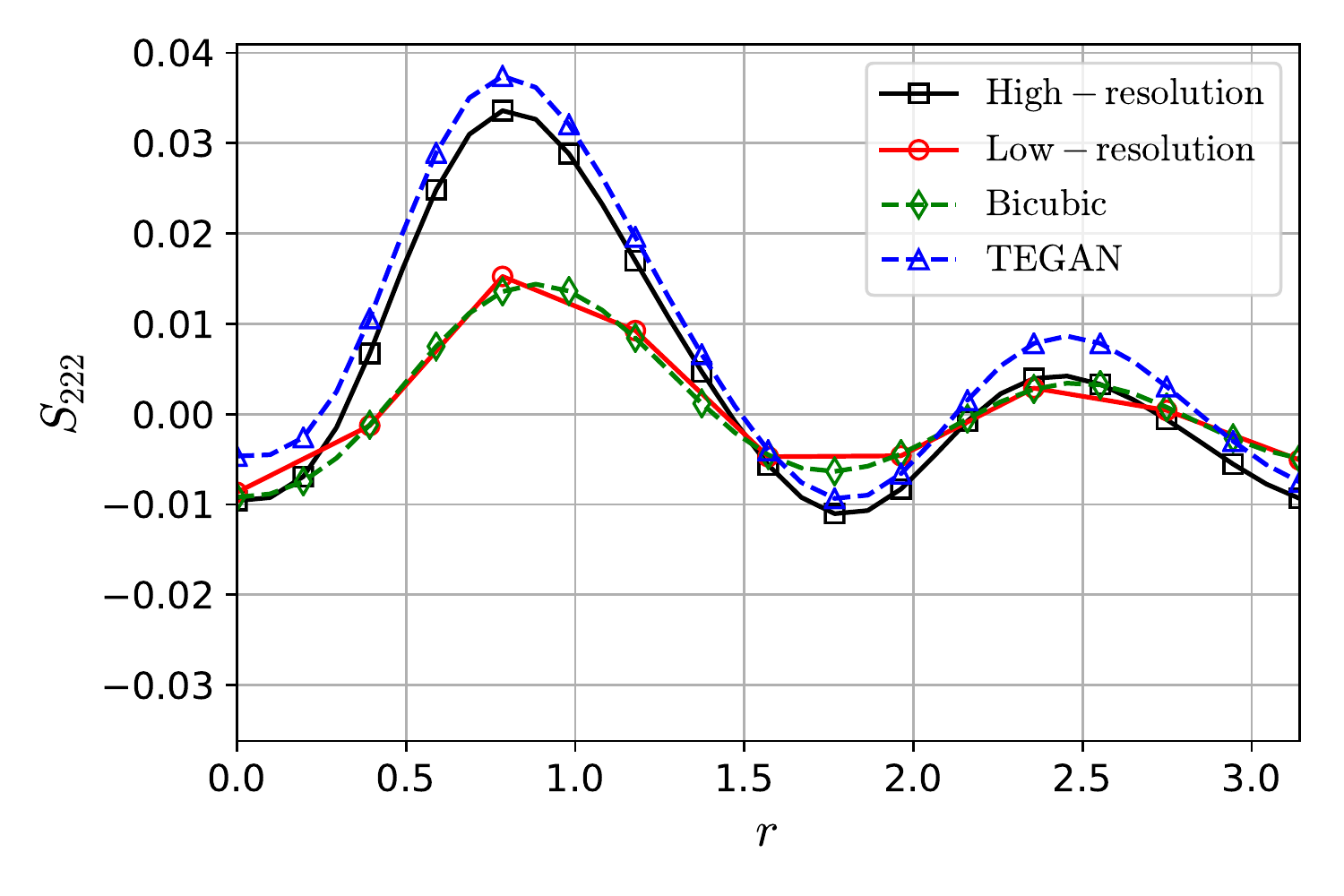}}
    \caption{Longitudinal (left) and transverse (right) two-point triple velocity correlations compared from the low resolution data (red circles), high resolution data (black squares), bicubic interpolation (green diamonds) and TEGAN (blue triangles). \label{fig:TPC_S}}
\end{figure}

Two-point correlations represent the energy in a field at different length scales. However, turbulence is a dynamic and non-linear phenomenon and many of the complexities of turbulent flows arise from the interaction between different length scales. The Karman--Howarth equation describes these energetic interactions across length scales in a statistical sense and is given by
\begin{equation}
    \frac{\partial}{\partial t}\left( {u^\prime}^2 f \right) - \frac{{u^\prime}^3}{r^4} \frac{\partial}{\partial r}\left(r^4 \bar{k}\right) = \frac{2\nu {u^\prime}^2}{r^4} \frac{\partial}{\partial r} \left( r^4 \frac{f}{\partial r} \right)
\end{equation}
where $u^\prime$ is the RMS velocity fluctuation, ${u^\prime}^2 f\left(r,t\right) = D_{11}\left(r\bm{e}_1,t\right)$ is the longitudinal two-point correlation function and ${u^\prime}^3 \bar{k}\left(r,t\right) = S_{111}\left(r\bm{e}_1,t\right)$ is the longitudinal two-point triple velocity correlation function. Thus, transfer of energy from large to small scales of turbulence is represented by the $\bar{k}$ term in the Karman-Howarth equation above and signifies the importance of the longitudinal two-point triple velocity correlation function $S_{111}\left(r\bm{e}_1,t\right)$ in the dynamic energy cascade process.

Figure~\ref{fig:TPC_S} shows the longitudinal and transverse two-point triple velocity correlation functions $S_{111}\left(r\bm{e}_1,t\right)$ and $S_{222}\left(r\bm{e}_1,t\right)$ respectively. Here again, simple bicubic interpolation only acts to interpolate the low resolution triple correlations in $r$ and not recover the true high resolution data. The TEGAN model is able to represent the true triple velocity correlations much better and is with a maximum error being a 15\% overprediction of the longitudinal triple correlations at $r = 0$.  Third order statistics are typically much harder to reproduce than second order statistics for many enrichment methods and the ability of a generative model to represent these is significant to many applications sensitive to energy transfer mechanisms.

\begin{figure}
    \centering
        \subfloat{\includegraphics[width=0.5\textwidth]{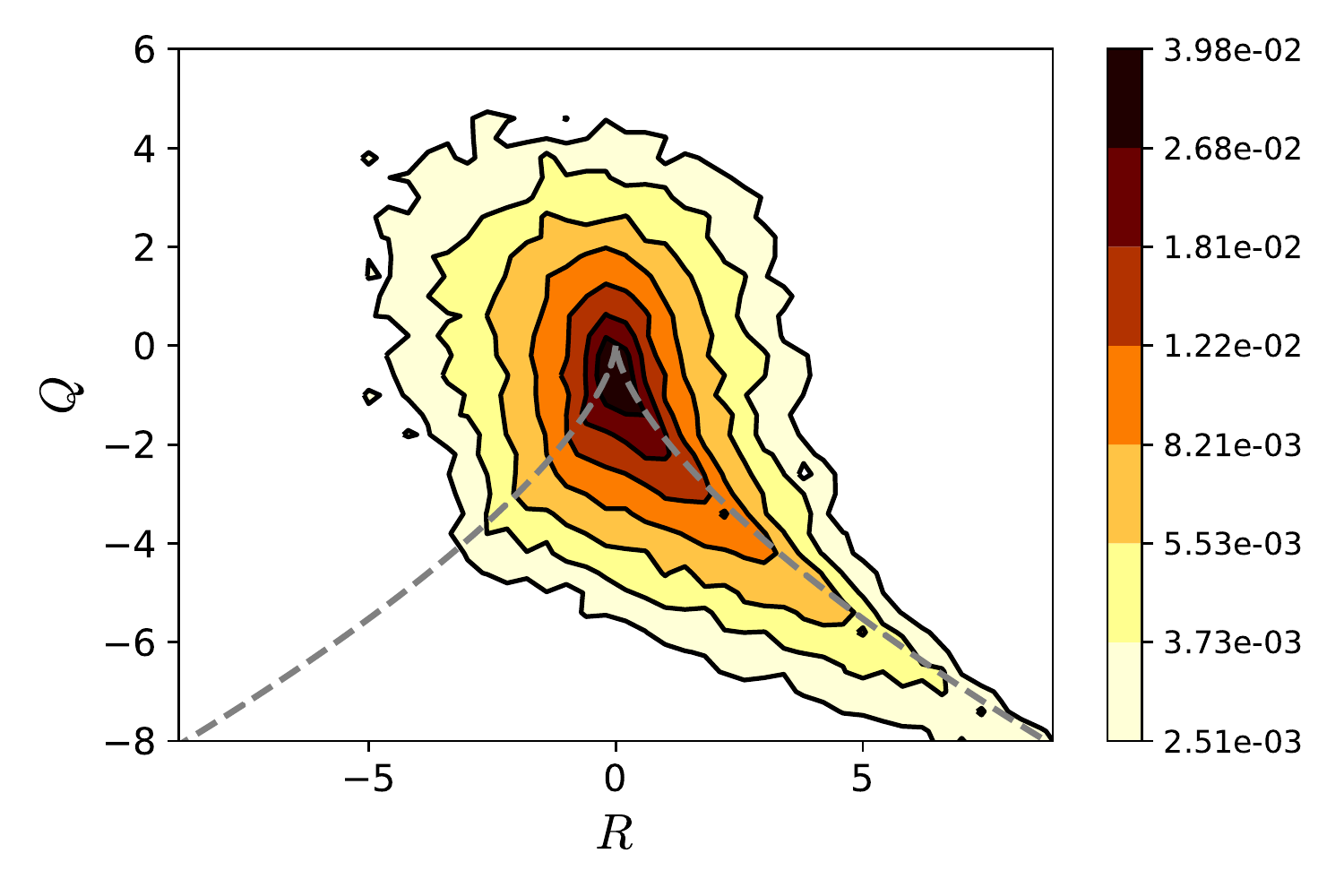}}
        \subfloat{\includegraphics[width=0.5\textwidth]{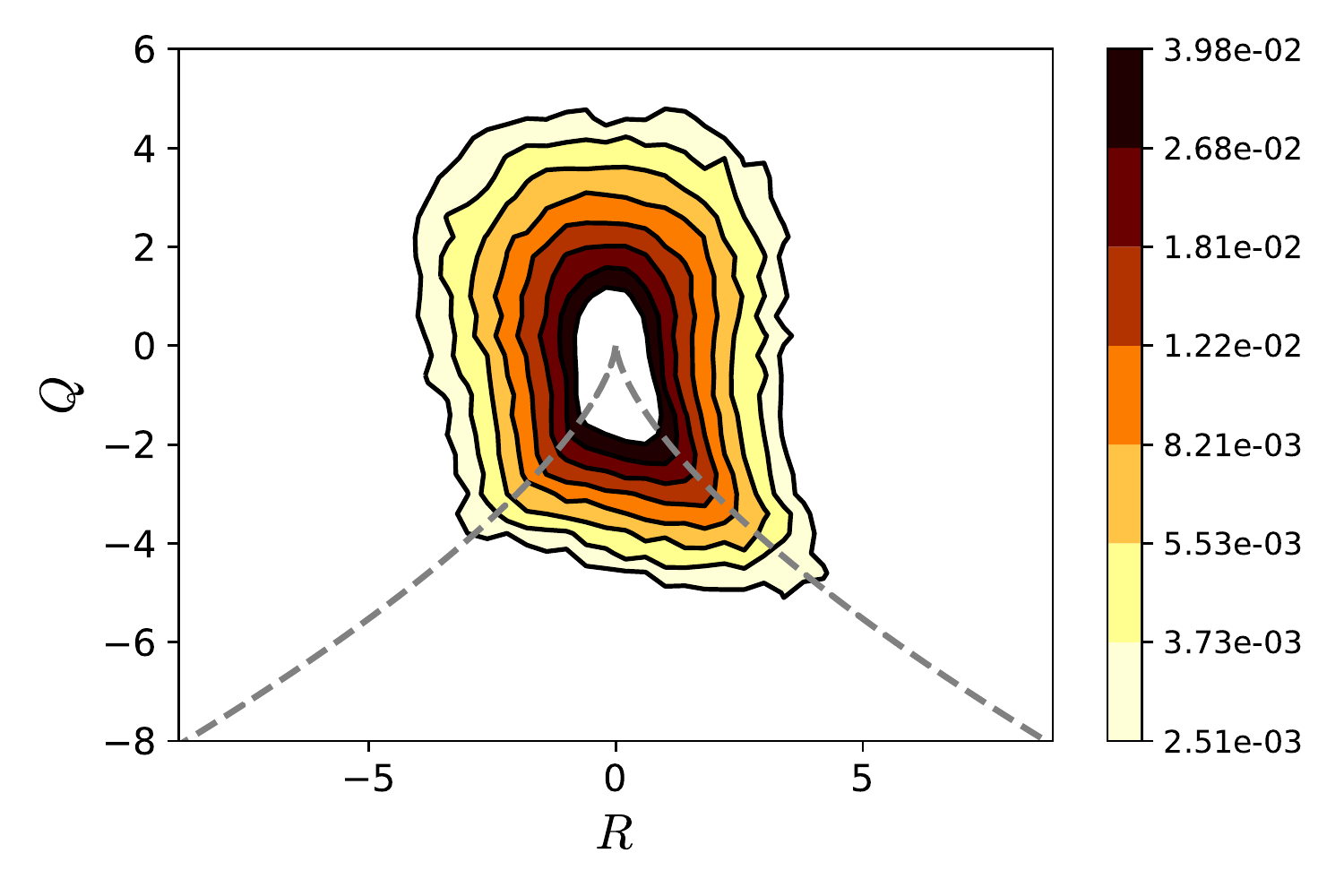}}\\
        \subfloat{\includegraphics[width=0.5\textwidth]{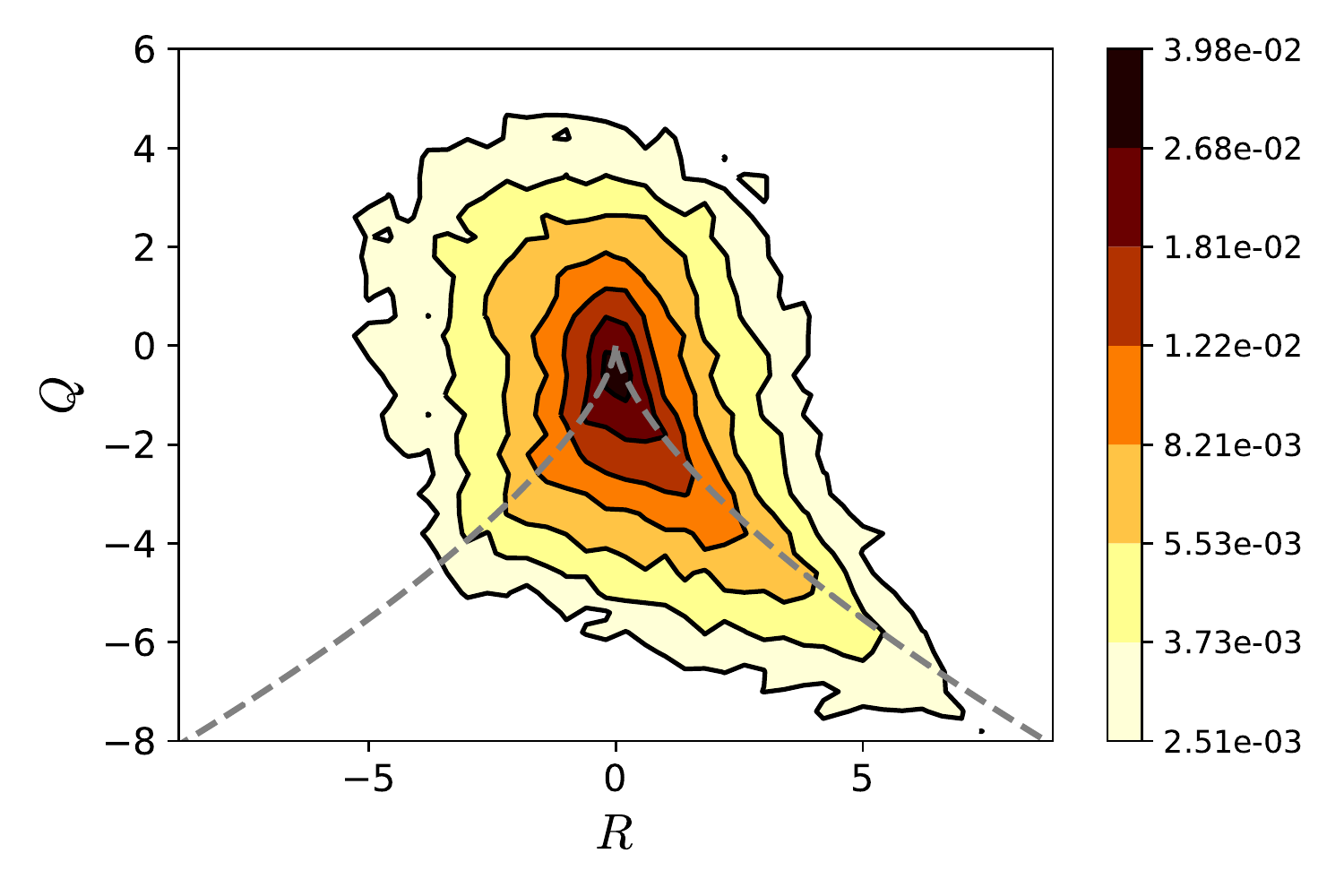}}
        \subfloat{\includegraphics[width=0.5\textwidth]{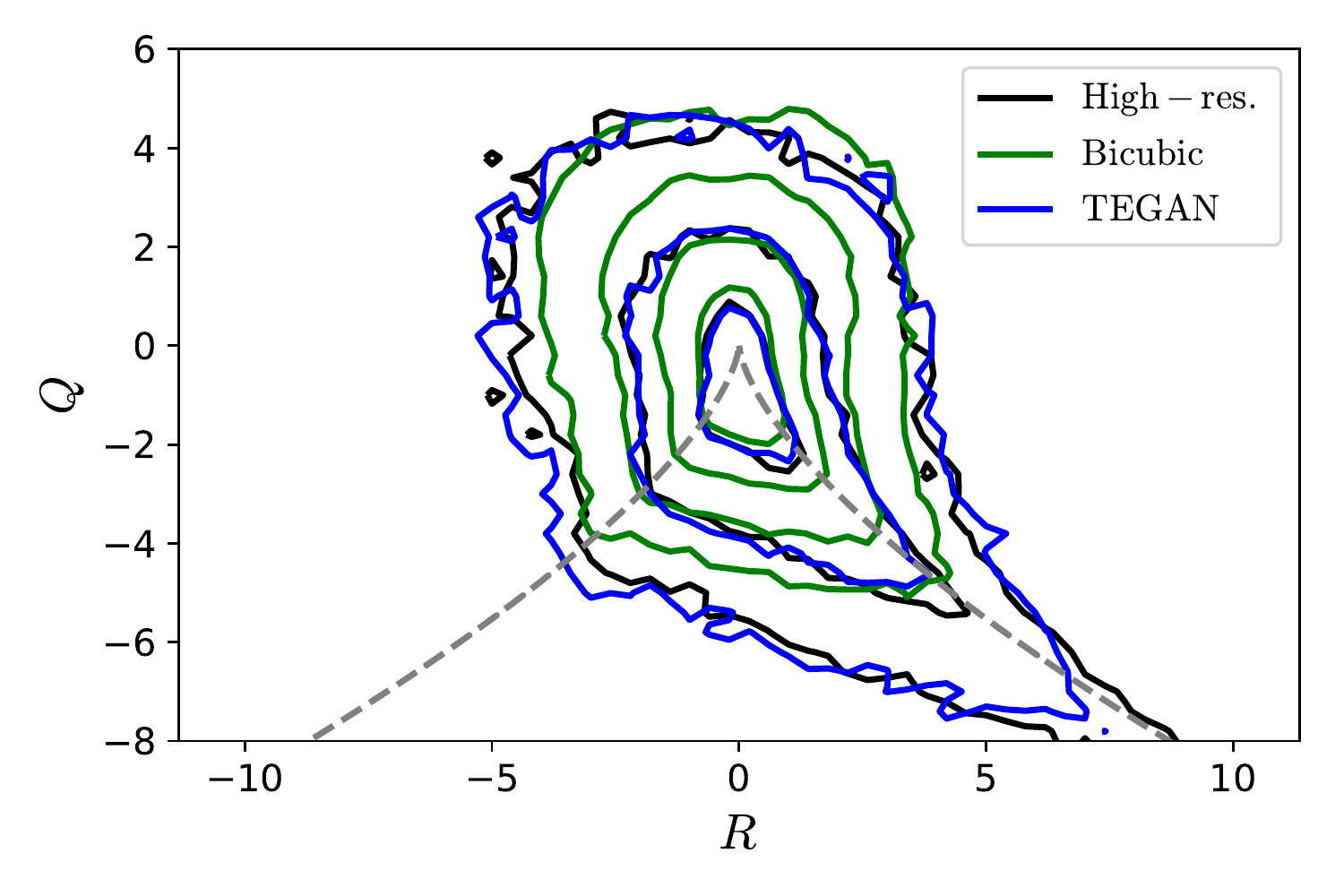}}
    \caption{$Q-R$ diagrams for the high resolution data (top left), bicubic interpolation (top right), TEGAN (bottom left) and contours comparing the three methods (bottom right). The grey dashed lines represent the restricted Euler solution to the $Q$ and $R$ invariants of the velocity gradient tensor \cite{cantwell1993behavior}. \label{fig:QR_plots}}
\end{figure}

The $Q-R$ diagram was introduced by \citet{chong1990general} and \citet{cantwell1993behavior}. It describes the PDF of the invariants of the velocity gradient tensor and thereby serves as a comparison of the flow morphology between the data generated by TEGAN, the high resolution data and bicubic interpolation. Figure~\ref{fig:QR_plots} plots the $Q-R$ diagrams for the three methods and a comparison between them.  The restricted Euler solution (plotted in grey dashed lines) predicts that the flow evolves to the bottom right of the $Q-R$ diagram. We see that to be the case in the high resolution data. The structure of the $Q-R$ diagram from the TEGAN data is also very similar to the high resolution data with some differences in the bottom right of the plot that indicates differences in the very fine scales. Bicubic interpolation, as one would expect, cannot recover the right structure of the $Q-R$ diagram and is very different from the high resolution data.

\section{Conclusion/Future Work }
In this work, we present two models, TEGAN and TEResNet, to enrich low resolution turbulence data and try to recover DNS level quality by adding fine scales features. We show the effect of different losses and the impact of physics-based losses for improved performance of the networks. These physics-based losses essentially act as a regularizer that brings the model onto a manifold of those that are physically realizable. While both TEResNet and TEGAN outperform traditional bicubic interpolation, TEGAN captures the physics better than TEResNet and has better generalization to the test dataset.

The enriched data from TEGAN was then compared to the high resolution DNS data and contrasted against data from simple bicubic interpolation. TEGAN is able to enrich the low resolution fields and recover a large portion of the missing finer scales. This is shown through a comparison of the energy spectrum and of longitudinal and transverse two-point correlation functions. TEGAN is also able to represent the energy dynamics across scales as represented by the third order structure functions. Finally, we compare the $Q-R$ diagrams from the different method and show that TEGAN is able to represent the flow morphology well.

The current TEGAN architecture and training can be further improved in the future by the implementation of gradient penalty based Wasserstein GAN (WGAN-GP) \cite{gulrajani2017improved} for overcoming few of the shortcomings of the traditional GANs mainly related to stabilized learning. Also, using wider distributions of data and tasking the discriminator with physics-based classification along with discrimination can be explored for better performance of the TEGAN in the future.

\section*{Acknowledgements}
We gratefully acknowledge the help from Dr. Aditya S. Ghate in setting up the forced isotropic turbulence simulation problem in the PadeOps\footnote{\url{https://github.com/FPAL-Stanford-University/PadeOps}} code used to generate the simulation data.

\begin{appendices}

\section{Low-pass filter}\label{app:low_pass_filter}

\begin{figure}[ht!]
\centering
\includegraphics[width=0.5\textwidth]{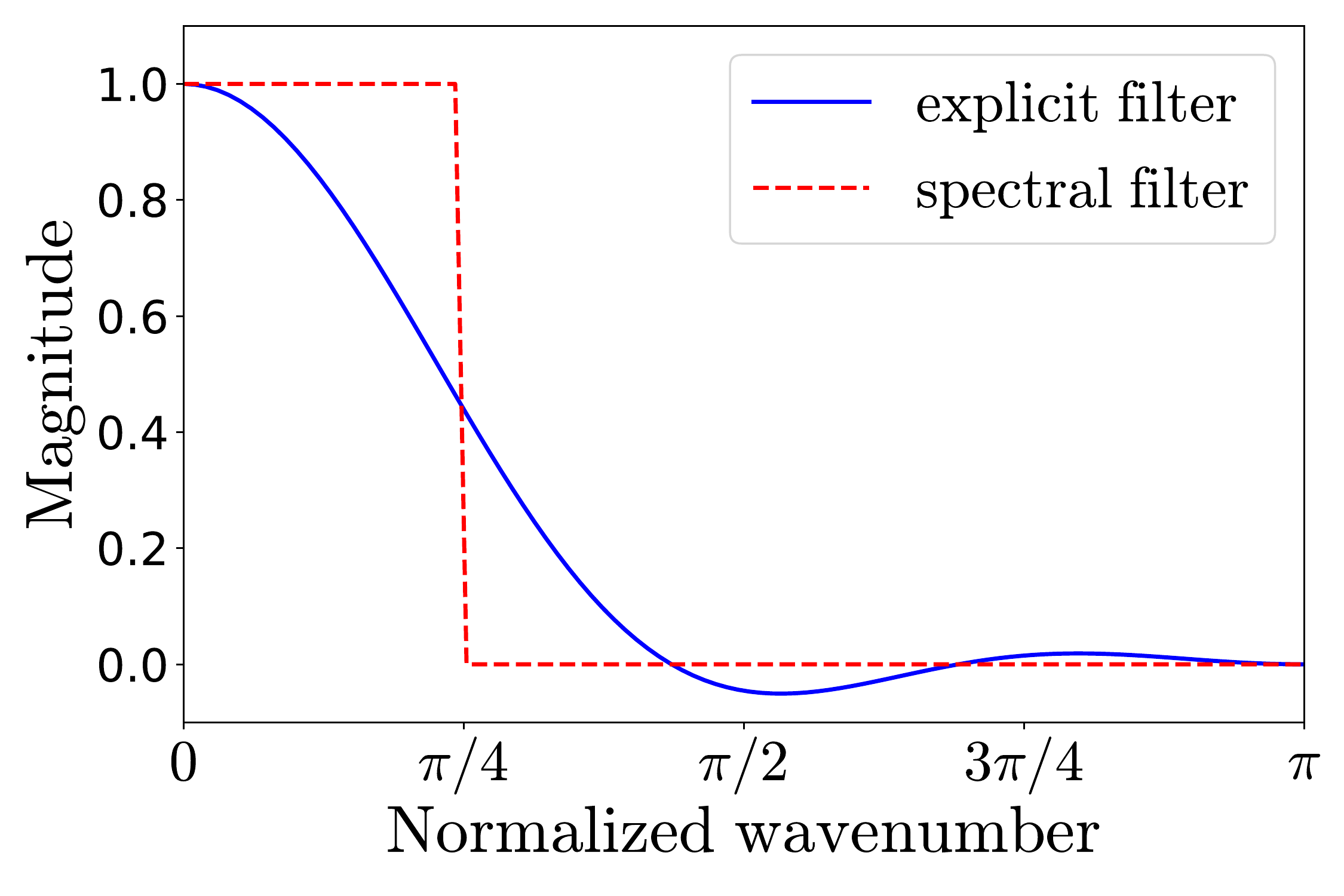}
\caption{Transfer function of the explicit filter.}
\label{fig:filter}
\end{figure}

An explicit low-pass filter with compact support is applied to the high resolution data before it is downsampled $4\times$ to low resolution data for training. The formula for filtering a variable $u$ at grid point $x_j = j \Delta x$, where $\Delta x$ is the grid spacing, on a uniform grid is given by:
\begin{equation}
\begin{split}
    \bar{u}_j &= 0.22723004 u_j + 0.20002636 \left( u_{j-1} + u_{j+1} \right) + 0.13638498 \left( u_{j-2} + u_{j+2} \right) \\
    &\quad + 0.04997364 \left( u_{j-3} + u_{j+3} \right),
\end{split}
\end{equation}
where $\bar{u}_j$ is the filtered variable. The transfer function of the low-pass filter is shown in Figure~\ref{fig:filter} and is compared with a spectral filter.

\end{appendices}

\bibliographystyle{unsrtnat}
\bibliography{refs}

\end{document}